# Evidence of coherence in strong-field electron photoemission from a semiconductor

Marie Froidevaux[1,*], Ludovic Douillard[2], Willem Boutu[3], Milutin Kovacev[4], Philippe Zeitoun[1] and Hamed Merdji[1,**]

[1]Laboratoire d'Optique Appliquée, CNRS, Ecole Polytechnique, ENSTA Paris, Institut Polytechnique de Paris, Palaiseau, France
[2]Université Paris-Saclay, CEA, CNRS, SPEC, 91191, Gif-sur-Yvette, France
[3]Université Paris-Saclay, CEA, CNRS, LIDYL, 91191, Gif-sur-Yvette, France
[4]Leibniz Universität Hannover, Institut für Quantenoptik, Welfengarten 1, D-30167, Hannover, Germany
*marie.froidevaux@ensta-paris.fr
**hamed.merdji@polytechnique.edu

**ABSTRACT**
Strong-field quantum electronics is emerging as a potential candidate in information processing but still coherence vs decoherence is a primary concern of the concept. Strong-field coherent processes in band gap materials have led during the last decade to the emergence of high harmonic generation in semiconductors, petahertz electronics, or strong-field quantum states. However, the coherent behavior of the sub-optical cycle-driven electrons has never been directly observed. We report here on the experimental evidence of coherent ultrashort emission of hot electrons from a nanostructured semiconductor. Our method uses sub-wavelength electric field enhancement to localize the electron emission within a nanometer-scale spot. We found similarities with the electron emission from metallic nanotips in the strong-field regime, a topic that has opened a vast domain of applications during the last decade. The electron spectra display both odd and even harmonic orders of the driving femtosecond laser frequency, a signature of the coherent nature of the electron emission and their attosecond timing. Our findings complete our knowledge of phenomena governing coherent strong-field processes in semiconductors and open perspectives for the generation of future quantum devices operating in the strong-field regime.

Advances in semiconductor science and ultrafast photonics offer a wide range of options for exploring new physics, particularly in extreme laser-matter interaction regime. Among these, the strong-field regime is an emerging research area with promising applications such as petahertz lightwave electronics[1–3], quantum information[4,5], or strong-field topology[6,7]. Strong-field-driven control of the electron state can be applied to store and process quantum information at unprecedented speeds, and in an unconventional manner compared to classical photonics. The use of various semiconductors such as low-dimension materials as well as functionalized nanostructures opens up new practical research directions to create compact and integrable quantum devices, fully controlled by laser light. We investigate here the coherence of bursts of hot electrons when submitted to an intense femtosecond laser. This electron emission can be accompanied by a coherent phenomenon discovered a decade ago in a zinc oxide semiconductor: high harmonic generation (HHG)[8]. The coherence of the harmonic photons has been practically applied in nanoscale coherent diffractive experiments[9,10] or has been used to create nanostructured topological beams[11]. However, the coherent nature of the electron emission phenomenologically linked to the high harmonic emission from a semiconductor has never been directly revealed. Most studies related to nanoscale emission of electrons deal with metals and specifically metallic tips[11–18]. For instance, Hommelhoff et al. reported on the attosecond control of electron emission from a tungsten tip[11]. Few studies are related to the ultrafast electron field emission in semiconductors. Borz et al. reported on the generation of a highly efficient thermal-damage-free source of hot electrons based on a single-crystal diamond needle[19]. All these studies share a characteristic feature, the presence of discrete energy peaks separated by one photon energy of the driving laser on the electron energy distribution curve (EDC), a signature of electron multiphoton-field-emission, well-known in above-threshold ionization (ATI) of atoms[20,21].

Here, we investigate the nature of the electron emission from a zinc oxide (ZnO) nanostructure irradiated by a femtosecond laser in the strong-field regime. The electron emission called above-threshold photoemission (ATP) shows characteristic discrete peaks separated by the driving laser photon energy corresponding to the number of photons absorbed by an electron with a characteristic intensity-dependent cutoff signature in the measured spectra. The elastic nature of the ATP peaks indicates that the process is driven by the strong laser field at the sub-optical cycle level.

**RESULTS**

**Nanoscale localization of the electron emission**

The strong-field regime is reached in the semiconductor using sub-wavelength focusing of the driving laser. The sketch of the experiment is shown in Figure 1a. A nanostructured ZnO crystal is irradiated from its base (rear side of the crystal) by a femtosecond tunable laser emitting in the 680-1080 nm spectral range (see Methods). The nanocone structuration produces a wave-guiding effect during the propagation of the light inside the structure to achieve a sub-wavelength enhancement of the electric field[9]. The semiconductor structure milled using a focused Ga-ion beam is shown from two perspectives in Fig.1b, c obtained by scanning electron microscopy (SEM). The electrons emitted from the truncated top shallow layer of the cone are collected by a photoemission electron microscope (PEEM) working in both spatial imaging and electron kinetic energy spectral modes (see Methods).

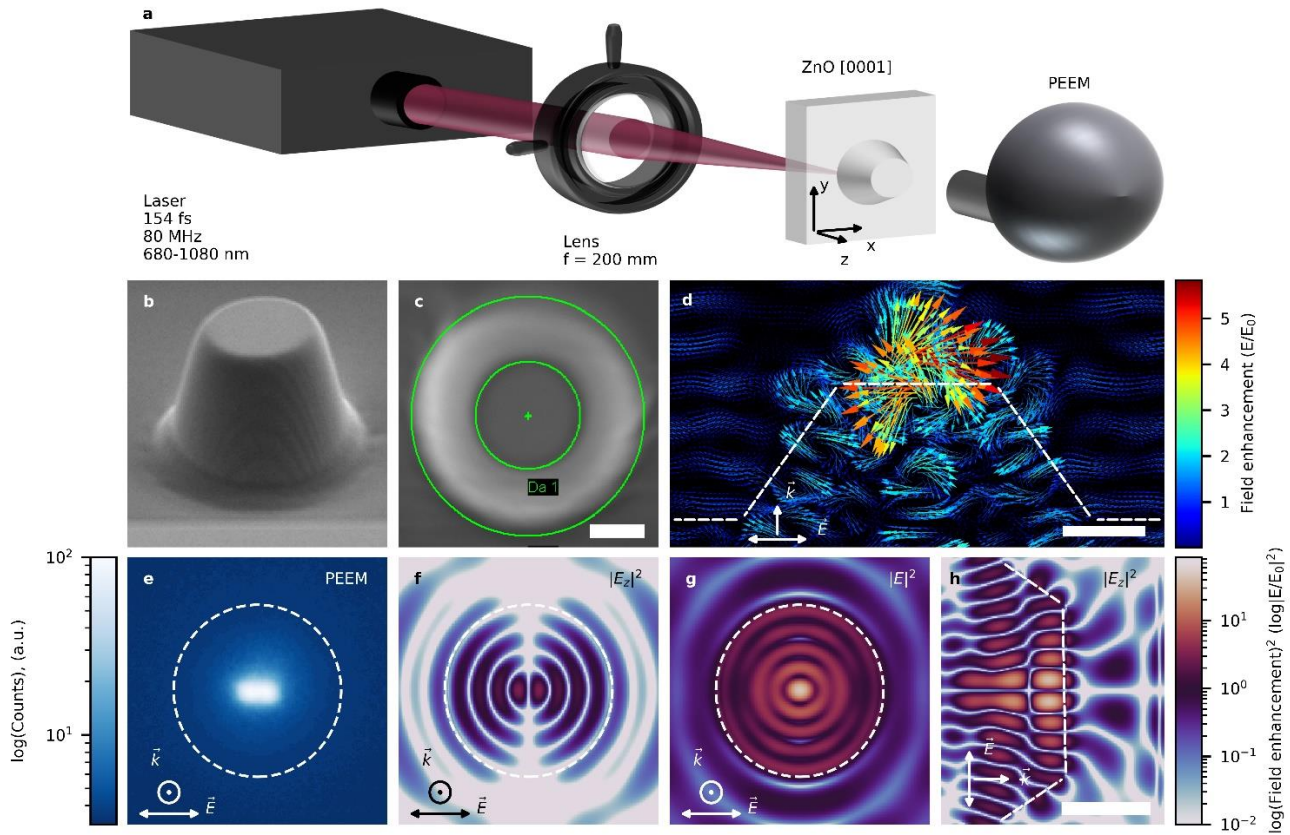

**Figure 1. Field emission from the nanostructured semiconductor**. **a**, Sketch of the setup: An intense IR fs laser beam is focused in a transmission geometry in a nanostructured cone etched on a 500 μm thick [0001] ZnO substrate. The emitted electrons are detected using a photoemission electron microscope. **b**, 60° view image of the truncated nanocone (SEM). **c**, Top view SEM image of the nanocone, white bar is 1 μm. **d**, FDTD simulations of local polarization directions indicated by color-coded arrows (the explicit color bar is displayed to the right). The white bar is 1 μm. **e**. Electron field emission from the top of the truncated nanocone measured by the PEEM (under 802 nm irradiation). **f**-**h**, Linear polarization FDTD simulations of the electric field distribution at optimized real truncated cone obtained with FIB at a laser wavelength of 802 nm. **f**, Square modulus of electric field *z*-component at the *xy*-exit plane, **g**, Square modulus of the total electric field at the *xy*-exit plane. **h**, Square modulus of electric field *z*-component along *xz*-cut. **f** and **h** – show the origin of the 2 elongated lobes in **e**. *E* and *k* are the electric field and the k-vector, respectively. The white bar in the lower right corner of figure **h** represents 1 μm and applies to all of the images **f**-**h**. The figures **f**-**h** share the same color of log scale 0,01 to 85.

Local enhancement of the electrical field amplitude by almost 2 orders of magnitude is predicted using Finite Difference Time Domain (FDTD) calculations (See Methods). Thus, the incident laser light field focused at a few 10 GW/cm² intensity regime rises locally up to few TW/cm² at the top exit surface of the ZnO structure on a sub-micrometer area. In short, the nanostructured cone is working as a sub-wavelength focusing lens. The FDTD simulations allow estimating the local field enhancement for the full spectral range of the laser source with any tailored polarization. The maximum achieved local electric field enhanced by the nanostructured cone is about few V/nm.

The experimental spatial distribution of the photo-emitted electrons shows a strong local field confinement (Fig.1**e**). The analysis of the spatial profile of the emitting electron spot displays an elongated two-lobe shape, 300x80 nm² in size (see Fig.1**d**). In fact, the emitted electrons come from an emitting zone located at the apex of the truncated cone, whose symmetry follows the distribution of the extracting *z*-component of the electric field for different polarization states of the laser light excitation as shown in Figs.1**f**. The lobes are separated by a 55 nm distance and illustrate the presence of a field nodal plane in agreement with the FDTD simulation.

**Semiconductor potential barrier bending**

The dimensionless Keldysh parameter $\gamma = \sqrt{\frac{W}{2U_p}}$ is introduced to investigate the main mechanism responsible for the electron emission process, where *W* is the work function, i.e. the minimal energy required to remove an electron from a solid, $U_p = e^2E^2/4m_e\omega^2$ is the ponderomotive energy, *e* is the electron charge, *E* is the local electric field amplitude, $\omega$ is the laser angular frequency, $m_e$ is the electron mass. For the local field intensities considered here (up to TW/cm², γ~1), the electron emission occurs in a mixed regime where multiphoton absorption and field emission take place simultaneously[18,22,23]. Thus the electron emission regime is treated within the framework of the field emission phenomenon similarly as it occurs at a metal surface submitted to a high static electric field, i.e. a Fowler-Nordheim process. The electron signal comes from the tunneling of electrons across a potential energy barrier. For a maximal local intensity of 1 TW/cm² the potential barrier's reduction is 3.31 eV. Therefore, the effective potential barrier is low, less than 0.01 eV. The estimated electron mean free path in ZnO at 300 K is around 3 nanometers which is about the thickness of the potential barrier obtained at the local laser intensity (see Fig.2**a**). From a physical point of view, the potential barrier can be dynamically affected by the applied electric field (see Fig.2**a**). Indeed, in the multi-cycle regime, the potential barrier can change from one optical cycle to the other during the pulse, giving rise to different parts of the curve, associated to electron emission below (tunneled) and above () the top of the surface barrier, respectively[24].

**Electrons kinetic energy ATP signatures**

Fig.2**b** presents the kinetic EDC of the electrons emitted at the cone's top hot spot. The peaks to the right from the multiphoton photoemission (MPPE) peaks are spaced by approximately one laser photon energy.

This is a clear indication of the ATP nature of the photoemission process[11,13,18]. Here, the dynamic field emission phenomenon is assisted by a multiphoton absorption (see Fig.2a) and occurs on an attosecond time scale giving rise to petahertz electron currents[1–3]. Together with the discrete ATP peaks, the electron EDC of ZnO shows a high-energy plateau feature followed by a sharp drop-off of the electron signal. In metals, such a plateau is a solid hallmark of an electron scattering process at the surface of the solid. The rescattering phenomenon corresponds to the return of the field-emitted electron wave packet to the surface following the inversion of the electric field during one optical cycle. This phenomenon determines the maximum achievable kinetic energy, i.e. the high cutoff energy of the electrons.

Figure 2d displays the kinetic energy distribution from a concatenation of partial spectra recorded at a wavelength of 784 nm and adjusted with a variable retarding potential (RP). The experimental kinetic energy electron curve Fig.2d can be visually divided into four parts tentatively attributed to different physical processes: (1) a pronounced MPPE reference peak at around 1.4 eV, (2) a multiphoton assisted field electron emission to the left of the reference peak, (3) a plateau to the right from the main MPPE peak (where 4 discrete secondary peaks are directly visible), and (4) a cutoff at high kinetic energy - a signature of strong-field emission and elastic scattering (ATP peaks). The presence of a plateau (3) and a high-energy cutoff (4) are similar to the observations of HHG spectra in gases and in band gap crystals. This illustrates the strong optical field electronic response of the ZnO semiconductor.

To further analyze the kinetic energy electron curve we performed a linear fit of the ATP peaks' positions below and above the MPPE separately. Up to 15 multiphoton ATP peaks are recovered as reported in Fig.2c, d (the resulting fit is shown as a grey dotted line in Fig 2d) with half of the peaks either below or above the reference photoemission peak (red solid line). The first peaks are clearly visible in the raw data, the others are partially smeared out. We observe in Fig.2b, d that the ATP peaks are partially shifted to lower energy which may have multiple origins. Krueger et al[13] reported a loss of the visibility of elastic emission peaks due to the carrier-envelope phase but this effect do not apply to our long pulses. In our case, to explain the smearing out of elastic scattering peaks we may consider a dynamical band gap reduction due to career injection at each optical cycle as reported by Schultze et al[25] that would lead to a shift of the photoelectrons energy during the multi-cycle pulse. Another possible origin of the smearing out and shifting of the elastic peaks can be attributed to inelastic processes such as electron-electron collisions, which are quite dominant around the band gap of ZnO[9]. Additionally, lattice modifications may occur when electrons tunnel from the valence band states to the conduction band states, which may lead to a band gap reduction. However, the lattice response to the laser pulse occurs on much longer time scales than electronic processes and is less likely to influence the ATP peak emission during the pulse. Finally, the ATP peak's energy broadening can be due to the Coulomb interaction during electron propagation towards the detector. The emitted electron cloud will push back the slower electrons and push forward the faster electrons providing respectively a loss and a gain of kinetic energy to the initial electron energy thus enlarging the ATP peaks[17].

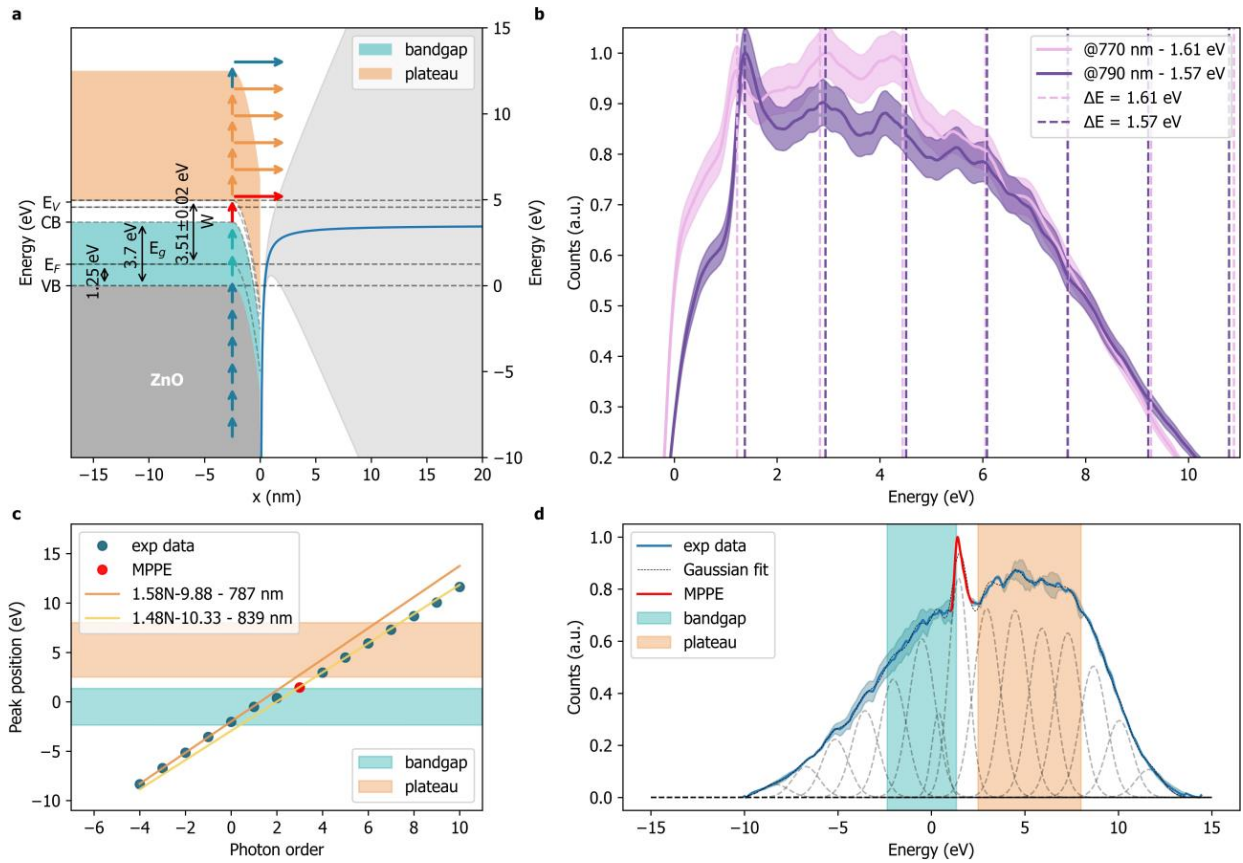

**Figure 2. Electron kinetic energy distribution. a**, Schematic of the ZnO energy diagram. The potential energy experienced by an electron is modelled as a rounded quasi-triangular barrier, i.e. a Schottky-Nordheim barrier model - blue line. The electric field varies with ±1.9 V/nm intensity range (grey-filled area). The color code of arrows and areas is as follows: red - electrons corresponding to the MPPE peak (see the text), orange - electrons escaping from above the CB and forming plateau in **d**, cyan - electrons escaping from the bandgap area virtual levels. VB - Valence band, CB - Conduction band, $E_F$ - Fermi level, W - work function. Arrows represent 1.58 eV (784 nm). **b**, Electronic energy curve at 770 nm (mauve) and 790 nm (violet) measured using a linear laser polarization at 0.48 and 0.46 TW/cm² laser intensity respectively. Light colors around curves indicate a 5% noise area. Distance between dashed cyan and violet lines correspond to 1 laser photon energy respectively at 770 nm (1.61 eV) and 790 nm (1.57 eV), starting from the MPPE. **c**, Positions of the discrete energy peaks extracted from the EDC curve in **d** by a least square fit of the data with multiple Gaussian distributions before the MPPE peak and along the plateau. **d**, Full electron EDC linearly polarised at 784 nm wavelength and 0.51 TW/cm² laser irradiance. Photons' energy is coherently summed up during the pulse duration and while the potential barrier varies within the grey area, electrons escape from the material and form the electron energy distribution curve. The solid blue line is the experimental data, the transparent blue area around the experimental data is the result of average, minimum, and maximum curves obtained with several RP positions. Dashed and dotted grey lines are the sum of multiple Gaussian fits and separate Gaussian distributions, respectively.

The attribution of the central peak to a coherent multiphoton absorption is based on two physical arguments. First, at low laser irradiance, there is no significant field effect and the electron EDC exhibits only one narrow single peak. Second, the work function of ZnO has been determined *in situ* by measuring its electron reflectivity curve[26–29].

The experimental value (3.51±0.20 eV) is in good agreement with the literature[30, 31]. The signal on the low energy side of the multiple photoemission peak is interpreted as the result of a field emission phenomenon assisted by photoelectric absorption. Yanagisawa et al[32] studied the effect of ultrafast field emission of electrons from metal tips in a regime of photo-assisted field emission. The field emission current is mainly determined by two factors: (i) the electron occupation number and (ii) the transmission probability across the surface potential barrier. In metals, the electron distribution obeys a Fermi-Dirac distribution function. Upon laser irradiation, the electron distribution is modified by single-electron excitation due to multiphoton absorption, resulting in a non-equilibrium distribution characterized by a step-like profile progressively smeared out by electron-electron scattering of characteristic timescale 100 fs (compared to the pulse duration of 154 fs). Although developed in the framework of metals, this interaction scheme should remain applicable to semiconductors provided that the band gap feature is taken into account. Therefore, one expects the multiphoton absorption process to suffer from the low number of electronic states within the band gap. However, the rising edge of the electronic signature does not show any trace of the material's band gap[33]. This observation points to intermediate virtual states (see Fig.2**c**, **d**) and supports the existence of strong electric field amplitudes within the ZnO nanostructure. Note that only the first four peaks after the direct (non-linear) photoemission peak were taken into consideration. The reason to do so is due to the higher energy spread and low peak intensity at large kinetic energy, which makes it difficult to determine their exact positions. An analysis of the peak amplitudes reveals the existence of 6 electronic emission orders (steps like features) corresponding to coherent multiphoton absorption events at energies below the MPPE peak, so originating deeply within the valence band. So at variance to the case of metals for which the excited electrons come mainly from the Fermi level, here the electron population of the full valence band participates in the electronic signal. Again, the signature of the valence band electrons suggests the presence of a strong electric field within the crystal boosted locally by the nanostructure geometry.

**Electrons cutoff energy**

The impact of the *local* optical field on the cutoff energy of the kinetic electron energy curve was studied. Electrons emitted by the ionization of atoms in metals in strong-field regime exhibit cutoff energy which classically scales with the ponderomotive energy as $10U_p$[34]. The analysis of strong-field photoemission on dielectric nanospheres by Seiffert et al[35] revealed an effect of a short-range trapping field and an extended cutoff of $14.5U_p$ in case of energy gained due to backscattering and $9U_p$ due to recollision. Here, we measure cutoff energy that scales linearly with the local laser irradiance (see Fig. 3) due to the geometrical effect of the nanostructured cone. The cutoff energy was taken at 10% of the maximum electron count rate. The cutoff is corrected from the locally enhanced intensity by a parameter $\beta$ and its extension is then given by $14.5\beta U_p$. We apply to the ponderomotive energy an explicit local electric field enhancement factor $\alpha$, $U_p = \alpha^2 U_p^0$, where $U_p^0$ is the initial laser intensity, $\alpha = |E/E_0|$ is the total electric field enhancement factor, obtained from FDTD simulations (as electrons gain their energy in the 3D electric field) and is equal to 5.32. A linear fit to cutoff energy $E_{cutoff} \sim 14.5\beta U_p$ yields $\beta$ values of $17.4\pm2.4$ (see Fig. 3).

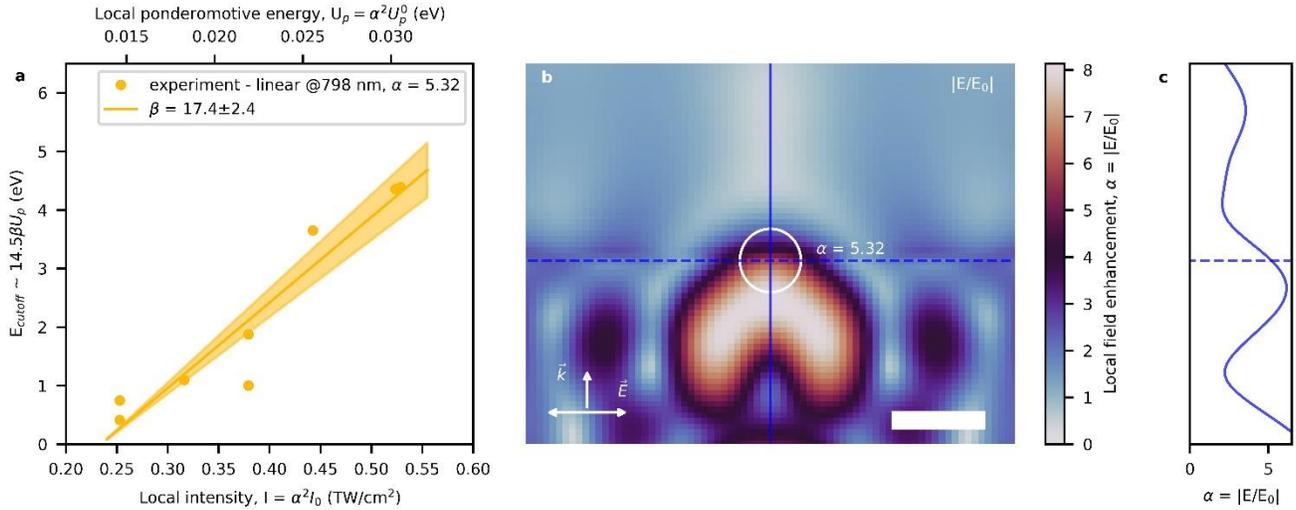

**Figure 3. Cutoff energy of the EDC. a**, EDC for different irradiances at the surface of the structure obtained experimentally with PEEM. Light area around curves - 10% error. **b**, Zoom in to the square modulus of the total electric field along *xz*-cut. The blue dashed line shows the exit surface of the nanostructure and the white circle shows where the parameter $\alpha$ was estimated with FDTD simulations. The blue solid line traces the local enhancement of the electric field in **c**. The white scale bar represents 300 nm. **c**, Local enhancement of the total electric field along the laser field propagation. The blue dashed and full lines have the same significance as in **b**.

High energy cutoff, up to 6 times the band gap of ZnO, is a result of (1) the rescattering of previously tunnel-ionized electrons in the enhanced near field in the vicinity of the surface and (2) the trapping potential produced by residual ions and other free electrons again in the vicinity of the surface, similar to studies in dielectric nanoparticles by Zherebtsov et al.[36]. Here, we suppose that this effect is due to the multi-cycle nature of the driving laser, thus electrons accelerate several times within the pulse, as well as in the trapping field in the vicinity of the nanostructured truncated cone which results in the enormous shift of the cutoff energy. Figure 2**b** shows a part of the electron EDC (~12 eV width) for two different wavelengths. Upon wavelength change, one observes the relative suppression of the MPPE peak in favor of the high-order multiphoton peaks which become more visible, a known phenomenon in ATP in metals[12]. This phenomenon is called onset on channel closing or threshold shifting and occurs in presence of the laser field, which shifts the continuum and increases the ionization potential. Thus, if the light-shifted potential barrier exceeds the energy of a given photon order, the corresponding peak will be suppressed due to the AC Stark effect.

## DISCUSSION

In summary, we have reported on the observation of ATP in a nanostructured semiconductor driven by a femtosecond laser pulse. The nanostructuration enhances the electric field at the exit plane of the ZnO nanocone by almost two orders of magnitude, thus reaching the strong-field regime necessary to achieve the hot electron emission. In comparison to electron emission from the metal tip, the proposed method of coherent pulsed hot-electron generation by local field enhancement is not affected by thermal load. The electron emission over time is therefore relatively constant and does not suffer from rapid material degradation. We have observed elastic electron emission peaks separated by the photon energy of the driving laser field, which is a direct signature of the coherent scattering process occurring at the sub-optical cycle level. We also observe a partial shift of the ATP peak to lower energy, which can be attributed to a band gap reduction during the pulse. In addition to the occurrence of ATP peaks, the coherent light-wave electron emission presents signatures similar to those encountered in metal nanotips operating in the strong-field regime such as a peak suppression at higher laser energies, the presence of plateau and an energy cutoff. However, in semiconductors, electrons originate from the valence band. We show that the electron emission is significantly increased reaching a 6 eV cutoff for a local intensity of 1 TW/cm$^2$. The large cutoffs are explained by multi-cycle many-body effects from accumulated charge at the surface of the semiconductor. Similar to HHG, we expect higher ATP cutoff from high band gap semiconductors.

As a perspective, our method can be potentially used to create an attosecond nanoscale source of electrons or create new directions in petahertz optoelectronics. On a more fundamental side, our findings can be applied to other strong-field effects such as recollision physics in semiconductors that can be used to probe ultrafast electron dynamics. Indeed, each ATP peak carries information accumulated during the electron wave packet travel before emission and can be used as a probe with possibly attosecond and atomic resolutions. Finally, the coherent nature of the photoelectron emission in semiconductors opens routes towards strong-field quantum information such as strong-field topology or valleytronics using complex laser waveforms[37]. Indeed, a fine analysis of the ATP peaks as readout observable may encode information about the system's quantum state.

## METHODS

### FDTD simulation

The simulations of the field enhancement have been performed with LUMERICAL Solutions (Ansys Canada Ltd). The finite difference frequency domain (FDTD) method allows for simulating the reliable optical field distribution in 3D taking into account the optical properties of the materials such as permittivity and optical field parameters: pulse duration, wavelength, polarization, and incidence angle. The local enhancement of the electromagnetic field in the nanostructured waveguide was optimized for the central wavelength of the source - 800 nm so that the highest local field intensity is achieved at the top of a truncated nanocone, where the electrons originate from. The ZnO optical parameters were taken from Bond et al[38]. The surrounding medium is vacuum. The structure is illuminated by a plane wave at 800 nm wavelength from the backside of the sample. Once the structure was optimised for the central wavelength the simulations were repeated for the range 750-850 nm to estimate the local electric field enhancement. Perfectly matching layers were used as the boundary conditions. A finite mesh of 15 nm was used in $x$, $y$, and $z$ directions. The calculation time for each structure is about 3 hours on a bi-processor Intel Haswell 10C E5-2650V3 (10 cores, 20 threads, max. frequency 3 GHz, Bus speed 9.6 GT/s QPI, 768 GB registered SDRAM (DDR4 2133, 68 GB/s bandpass)). The resulting geometry as well as the resulting enhancement are represented in Table 1.

| Polarisation | h (FDTD/SEM) | $\theta$ (FDTD/SEM) | $\alpha$ (FDTD) | $\beta$ (PEEM) |
|---|---|---|---|---|
| Linear | 1.6 μm/1.68 μm | $35^0$/$36.4^0$ | 5.32 | 17.4±2.4 |

**Table 1.** Nanostructured cones geometry optimized for linear polarization state, where $\alpha$ is the local electric field enhancement factor due to nanostructuration of the ZnO crystal, $\beta$ is the electrons' cutoff energy enhancement factor, $h$ is the height of the cone from bottom to top, $\theta$ is the angle between bottom diameter and the side of the cone and total field enhancement is the relative value showing the amplification of the local electric field.

### PEEM setup

The excitation light source used for PEEM measurements is a pulsed mode-locked Ti:Al$_2$O$_3$ laser system (Chameleon Ultra II, Coherent Inc.) delivering tunable photons in the 680-1080 nm spectral range (1.15-1.82 eV) (see Fig.1a). The pulse repetition rate is fixed to 80 MHz. Each pulse exhibits a profile of 154 fs duration. The light is focused at normal incidence ($0^0$) on the backside of the sample using a 200 mm focal length lens. The peak power densities delivered at the sample surface are in the sub-TW/cm$^2$. The photoemission electron microscope used is a commercial instrument (Elmitec GmbH LEEM/PEEM III) operating in an ultrahigh vacuum at a pressure in the low range of $10^{-10}$ mbar. Lateral spatial resolutions of 10 and 25 nm are routinely achieved in low-energy electron (LEEM and photoelectron emission (PEEM) imaging modes, respectively. The instrument operates in either total electron yield (no energy filtering) or spectrometric mode through the use of a hemispherical energy dispersive analyser (energy resolution 150 meV). The real-space images are recorded via a charge-coupled device (CCD) camera. In PEEM imaging mode, the measured brightness in a given image area is proportional to the electron emission from that area. Integrated PEEM signal is computed in two steps: (1) background subtraction, i.e., removing a regression plane, (2) integration of the pixel values over a representative image area. The collected electron emission reflects the optical near field at the surface of nanostructured object[26–29, 39].

**Sample preparation and characterisation**

The surface of a ZnO [0001] crystal is patterned with isolated truncated waveguide structures via focused ion beam (FIB) milling. Several structures optimised for linear polarisation were prepared. The ZnO work function is determined from the measure of the electron reflectivity in LEEM imaging mode[26–29, 39]. Absolute work function values are determined by the acquisition of a reference electron reflectivity curve on a freshly cleaved HOPG surface. Taking a reference value of 4.67 eV for the HOPG surface yields a value of 3.51±0.02 eV. For semiconductors, this energy is required to extract the electron from Fermi-level to vacuum.

**Acknowledgements**

PETACom (Petahertz Optoelectronics Communication) FET Open H2020 Grant No. 829153/ OPTOLogic (Optical Topologic Logic) FET Open H2020 Grant No. 899794/ TSAR project (Topological Solitons in Antiferroics) FET Open H2020 grant number 964931. We acknowledge the financial support from the French ANR through the grants PACHA (No. ANR- 17- CE30-0008-01), FLEX-UV (N° ANR-20-CE42-0013-02), and ATTOCOM grants (ANR-21-CE30-0036-02). We acknowledge the financial support from the French ASTRE program through the "NanoLight" grant. The authors thank Jocelyne Leroy (CEA Saclay/DRF/IRAMIS/NIMBE) for XPS measurements of the ZnO.

**Author contributions statement**

**Author contributions**

H.M. conceived the idea and supervised the project. Simulations were performed by M.F. W.B. prepared the sample at FIB. The PEEM measurements were performed by L.D. Data analysis was performed by M.F., L.D., P.Z., M.K.. All authors discussed the results and contributed to the writing of the manuscript.